\documentclass[a4paper,12pt]{article}

\usepackage{graphics, epsfig}

\begin{document}

\author{C. Barrab\`es\thanks{E-mail : barrabes@celfi.phys.univ-tours.fr}\\     
\small Laboratoire de Math\'ematiques et Physique Th\'eorique\\
\small  CNRS/UPRES-A 6083, Universit\'e F. Rabelais, 37200 TOURS, France\\
P.A. Hogan\thanks{E-mail : phogan@ollamh.ucd.ie}\\
\small Mathematical Physics Department\\
\small  National University of Ireland Dublin, Belfield, Dublin 4, Ireland}

\title{Bursts of Radiation and Recoil Effects in Electromagnetism 
and Gravitation}
\date{}
\maketitle

\begin{abstract}
The Maxwell field of a charge $e$ which experiences an impulsive 
acceleration or deceleration is constructed explicitly by 
subdividing Minkowskian space--time into two halves bounded by a 
future null--cone and then glueing the halves back together with 
appropriate matching conditions. The resulting retarded radiation 
can be viewed as instantaneous electromagnetic bremsstrahlung. If 
we similarly consider a spherically symmetric, moving gravitating mass, 
to experience an impulsive deceleration, as viewed by a 
distant observer, then this is accompanied by the emission 
of a light--like shell whose 
total energy measured by this observer is the same as the kinetic 
energy of the source before it stops. This phenomenon is a recoil 
effect which may be thought of 
as a limiting case of a Kinnersley rocket.\\


\end{abstract}
\thispagestyle{empty}
\newpage

\section{Introduction}\indent
Perhaps the simplest example in electrodynamics of an 
impulsive electromagnetic wave is produced when a charge 
$e$ receives an impulsive acceleration or deceleration. 
For example the charge $e$ might be moving rectilinearly 
with constant 3--velocity $v$ relative to the laboratory 
frame and is suddenly halted. This means that in Minkowskian 
space--time there is a sudden change in the direction of 
the 4--velocity of the charge. As a result a spherical 
impulsive electromagnetic wave is emitted. Although 
this physical situation has been known for some time 
\cite{P} no analytical description appears to have 
ever been given allowing calculations (such as those given 
in section 4 here) with the explicit 
Maxwell field of the charge. The construction of this 
Maxwell field is one of the objectives of the present 
paper. It involves subdividing Minkowskian space--time 
into two halves, each having as boundary the future null--
cone history ${\cal N}$ of the impulsive electromagnetic wave, 
and then glueing the halves back together whilst 
maintaining via {\it matching conditions} the invariance 
of a certain quantity. This construction is carried out 
in section 2 and a coordinate system is obtained in which 
the metric tensor of the re--attached space--times is 
continuous across the future null--cone ${\cal N}$. This enables 
us to see clearly that (i) the glueing ensures that the 
space--time remains flat on ${\cal N}$ and thus ${\cal N}$ cannot, for 
example,  be the history of a gravitational impulsive wave 
or shock wave and 
(ii) the electromagnetic field we construct satisfies 
Maxwell's vacuum field equations everywhere off the 
world--line of the charge $e$ and, in particular on the 
future null--cone ${\cal N}$ minus its vertex. We then define a 
measure of the intensity of the electromagnetic impulse 
wave and show that its dependence on the original 3--velocity 
$v$ of the charge is typical of the velocity dependence 
of the intensity of electromagnetic bremsstrahlung emitted 
in beta decay \cite{J}. We thus refer to the impulsive electromagnetic 
wave as instantaneous electromagnetic bremsstrahlung.

Having described instantaneous bremsstrahlung 
in electromagnetism one is naturally lead to
ask similar questions in the case of the gravitational
interaction: What happens when a mass receives an
impulsive acceleration or deceleration? In particular
will an impulsive gravitational wave accompany 
the abrupt change in motion of the mass?
Our geometrical approach to the electromagnetic case  
is perfectly adapted to this new situation. In section 3
we glue together two Schwarzschild 
space--times having as 
common boundary a future null--cone ${\cal N}$, using the same 
{\it matching conditions} as in the electromagnetic 
case. Also one of the Schwarzschild space--times involves 
a parameter $v$ which we show allows us to interpret its 
source as moving rectilinearly with 3--velocity $v$, 
compared to the source of the second Schwarzschild 
space--time, as measured by an observer at spatial 
infinity. Thus a distant observer sees a uniformly 
moving source come to a sudden halt. As a result a signal 
travels outward with history ${\cal N}$. We use the  
theory of light--like signals in general 
relativity developed by  Barrab\`es--Israel \cite{BI}
to show that this signal is a light--like 
shell (burst of neutrinos, for example). 
This emission of 
a light--like shell is similar to a rocket exhaust which 
carries sufficient energy to result in zero recoil velocity. 
We point out, following (3.20) below, the sense in which 
our model may be thought of as a limiting case of a 
Kinnersley rocket. Thus there is no gravitational 
radiation (i.e. it is not a spherical 
impulsive gravitational wave). Furthermore we demonstrate 
that the relative kinetic energy of the source, when 
travelling with 3--velocity $v$ measured by a distant 
observer, is converted into the energy of the light--like 
shell when the source suddenly stops. 
The physical interpretation of our model and the global structure
of spacetime will be discussed.

In section 
4 below we illustrate some calculations that can be carried out with 
the electromagnetic field obtained in section 2. 
These are based on the observations 
of Penrose \cite{P} that after a charge $e$ 
receives an impulsive acceleration the resulting spherical 
impulsive electromagnetic wave (a) approximates a plane impulsive 
electromagnetic wave in the far zone and (b) if it then experiences a 
head--on collision with a plane impulsive gravitational 
wave, the electromagnetic wave will back--scatter. This sort 
of phenomenon can be expected because pure unidirectional electromagnetic 
radiation must have a propagation direction in space--time 
which is shear--free (and geodesic) \cite{R}. However when the 
shear--free null geodesics intersect the history of a plane 
impulsive gravitational wave they must acquire shear \cite{P}. 
Hence the resulting electromagnetic field cannot be a pure unidirectional 
radiation field but must have another part to it (the back--
scattered radiation field). 
In section 4 the plane wave limit of the electromagnetic wave of  
section 2 is obtained and its subsequent head--on collision 
with a plane impulsive gravitational wave is considered. The 
electromagnetic field in the region between the waves after 
collision is derived by solving Maxwell's vacuum field 
equations with the appropriate boundary conditions. It 
is explicitly seen that the impulsive wave looses its plane 
character and the back--scattered radiation appears. In addition 
the electromagnetic field in this region after the collision 
becomes singular where light rays crossing the plane gravitational 
wave are focussed by that wave.

\setcounter{equation}{0}
\section{Electromagnetic Bremsstrahlung}\indent
Let $\left\{X^\mu\right\}$ with $\mu =1, 2, 3, 4$ be rectangular 
Cartesian coordinates and time in Minkowskian space--time, in 
terms of which the line--element reads (taking the velocity of 
light $c=1$)
\begin{equation} \label{2.1}
  ds^2=\left (dX^1\right )^2+\left (dX^2\right )^2+\left (
dX^3\right )^2-\left (dX^4\right )^2=\eta _{\mu\nu}dX^\mu\,dX^\nu\ .
\end{equation}
Let $X^\mu =x^\mu (u)$ be the parametric equations of the time--like 
world--line in Minkowskian space--time of a charge $e$ having 
$u$ as proper--time or arc length along it. The 4--velocity and 
4--acceleration of the charge have components $v^\mu (u)$ and 
$a^\mu (u)$ respectively (with $\eta _{\mu\nu}v^\mu \,v^\nu\equiv v_\mu\,v^\mu =-1$ 
and consequently with $v_\mu\,a^\mu =0$). Let $\left (X^\mu\right )$ be 
the coordinates of an event off the world--line of the charge 
and let $(x^\mu (u))$ be the coordinates of the event on the world--line
 of the charge where the past null--cone with vertex $\left (X^\mu\right )$ 
intersects the world--line. The retarded distance \cite{SRST} $r$ 
of $\left (X^\mu\right )$ from the world--line is given by
\begin{equation} \label{2.2}
r=-v_\mu\,\left (X^\mu-x^\mu (u)\right )\ .
\end{equation}
Here $X^\mu-x^\mu (u)$ is null and $r\geq 0$ with equality 
if and only if $\left (X^\mu\right )$ coincides with $\left (x^\mu (u)\right )$. 
The Li\'enard--Wiechert field of the charge evaluated 
at $\left (X^\mu\right )$ is given by the Maxwell tensor
\begin{equation} \label{2.3}
F_{\mu\nu}\left (X\right )=\frac{N_{\mu\nu}}{r}+\frac{III_{\mu\nu}}{r^2}\ ,
\end{equation}
with
\begin{eqnarray} \label{2.4}
N_{\mu\nu} & = & 2\,e\,q_{[\mu}\,k_{\nu]}\ , \\
III_{\mu\nu} & = & 2\,e\,v_{[\mu}\,k_{\nu]}\ .
\end{eqnarray}
Here square brackets denote skew--symmetrisation, $k^\mu=r^{-1}
\left (X^\mu-x^\mu (u)\right )$ so that $k^\mu$ is null and, by 
(\ref{2.2}), $v_\mu\,k^\mu =-1$. Also
\begin{equation} \label{2.5}
q^\mu =a^\mu +(a^\nu\,k_\nu )\,v^\mu\ ,
\end{equation}
and this is a space--like 4--vector orthogonal to $k^\mu$. The 
skew--symmetric tensor $N_{\mu\nu}$ in (\ref{2.4}) is Petrov type 
N with degenerate principal null direction $k^\mu$ and thus the 
leading term in the electromagnetic field of the charge (\ref{2.3}) 
describes the radiation part of the field. The presence of this 
term is due entirely to the acceleration $a^\mu$ of the charge. 
Suppose now that at $u=0$ (say) on the world--line of the charge, 
the charge receives an impulsive acceleration (there is sudden change 
in the 4--velocity of the charge at $u=0$ leading to a Dirac 
delta function $\delta (u)$ in the 4--acceleration of the 
charge) then one would expect the resulting retarded radiation, 
described by $N_{\mu\nu}$ above, to take the form of an impulsive 
electromagnetic wave with history the future null--cone $u=0$ 
and with profile $\delta (u)$. However this information cannot 
readily be extracted from (\ref{2.3}) and (\ref{2.4})--(2.5) because, 
among other things, $r$ in (\ref{2.2}) is now not defined at 
$u=0$ since there is no unique tangent to the world--line of the 
charge at $u=0$. To see this clearly let us take the vertex of 
the future null--cone ${\cal N}(u=0)$ to be the origin of the 
coordinates $\left\{X^\mu\right\}$ (and so $x^\mu (0)=0$). Thus if 
$P\left (X^\mu\right )$ is an event on ${\cal N}$ then $\left (X^\mu\right )$ 
satisfies 
\begin{equation} \label{2.7}
\left (X^1\right )^2+\left (X^2\right )^2+\left (X^3\right )^2
-\left (X^4\right )^2=0\ ,\qquad X^4>0\ .
\end{equation}
This is given in a useful parametric form by
\begin{equation} \label{2.8}
X^1+iX^2=\frac{\sqrt{2}\zeta\,R_0}{1+\frac{1}{2}\zeta\,\bar\zeta}\ ,X^3=
\left (\frac{1-\frac{1}{2}\zeta\,\bar\zeta}
{1+\frac{1}{2}\zeta\,\bar\zeta}\right )R_0\ ,X^4=R_0\ ,
\end{equation}
with $\zeta$ a complex variable with complex conjugate $\bar\zeta$ 
and $R_0$ a real variable. Putting $p_0=1+\frac{1}{2}\zeta\bar\zeta$ 
and substituting (\ref{2.8}) into the Minkowskian line--element 
(\ref{2.1}) yields the induced line--element on ${\cal N}$
\begin{equation} \label{2.9}
ds^2=2R^2_0p^{-2}_0d\zeta\,d\bar\zeta\ .
\end{equation}

\begin{figure}[!hb]
\center
\input{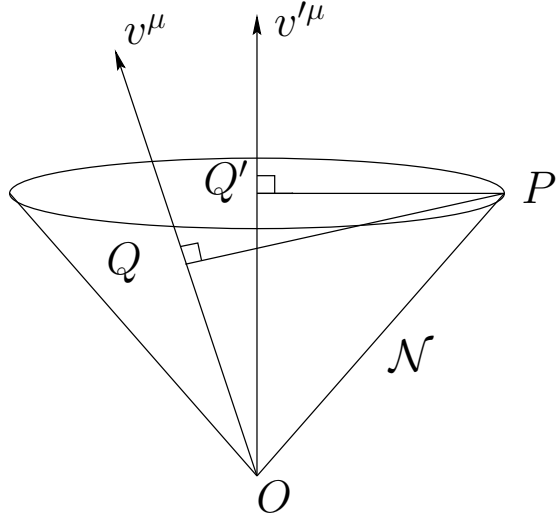}
\caption{\label{1}
The future null--cone ${\cal N}(u=0)$ with $QP=r$ and 
$Q'P=r_+$. The region to the future (past) of the null-cone
corresponds to $u>0\, (u<0)$.}
\end{figure}

%\centerline{----------------------------------------------------------------}
%\vskip 1truepc
%\centerline{Figure 1}
%\vskip 1truepc
%\centerline{-----------------------------------------------------------------}
%\vskip 1truepc\noindent
In figure 1, $QP$ is orthogonal to $v^\mu$ and $v_\mu v^\mu =-1$ while 
$Q'P$ is orthogonal to $v'^\mu$ and $v'_\mu v'^\mu =-1$. Then the 
retarded distances $r=QP$ and $r_+=Q'P$ are given by (using 
(\ref{2.2}) with $x^\mu (u)=x^\mu (0)=0$)
\begin{equation} \label{2.10}
QP=r=-v^1X^1-v^2X^2-v^3X^3+v^4X^4\ ,
\end{equation}
and 
\begin{equation} \label{2.11}
Q'P=r_+=-v'^1X^1-v'^2X^2-v'^3X^3+v'^4X^4\ .
\end{equation}
Substituting from (\ref{2.8}) into these we obtain
\begin{equation} \label{2.12}
r=R_0p\,p_0^{-1}\qquad {\rm and}\qquad r_+=R_0p_+\,p_0^{-1}\ ,
\end{equation}
with
\begin{equation} \label{2.13}
p=\frac{1}{2}\zeta\,\bar\zeta\,(v^4+v^3)-\frac{1}{\sqrt{2}}
(v^1-iv^2)\,\zeta -\frac{1}{\sqrt{2}}(v^1+iv^2)\,\bar\zeta 
+v^4-v^3\ ,
\end{equation}
and $p_+$ is the same function as $p$ but with $v^\mu$ in $p$ 
replaced by $v'^\mu$. Hence we see from (\ref{2.12}) that 
we have the {\it invariant statement}
\begin{equation} \label{2.14}
r_+\,p_+^{-1}=r\,p^{-1}\ ,
\end{equation}
and the line--element (\ref{2.9}) reads
\begin{equation} \label{2.15}
ds^2=2r^2p^{-2}d\zeta\,d\bar\zeta\ .
\end{equation}
Thus (\ref{2.14}) is a change of affine parameter along the 
generator $\zeta ={\rm constant}$ of ${\cal N}$ (OP in figure 1) 
which (a) leaves the vertex of ${\cal N}$ fixed and (b) leaves 
the induced metric on ${\cal N}$ invariant.

We now consider the following type of motion of the charge $e$: In 
the frame of reference with respect to which the coordinates 
$\left\{X^\mu\right\}$ are measured (the laboratory frame) the 
charge moves with uniform 3--velocity $v$ in the $X^3$--direction 
when $u<0$ (i.e. to the past of the null--cone ${\cal N}$). When $u>0$ 
(to the future of the null--cone ${\cal N}$) the charge is taken to be at rest 
in this frame. Thus for $u<0$ the world--line of the charge is 
the time--like geodesic with unit tangent (4--velocity)
\begin{equation} \label{2.16}
v^\mu =(0, 0, \gamma\,v, \gamma )\ ,
\end{equation}
with $\gamma =(1-v^2)^{-\frac{1}{2}}$. For $u>0$ the world--line 
of the charge is the time--like geodesic with unit tangent
\begin{equation} \label{2.17}
v'^\mu  =(0, 0, 0, 1)\ .
\end{equation}
The line--element of Minkowskian space--time in the region 
$u<0$ to the past of the null--cone ${\cal N}(u=0)$ and 
which we denote by ${\cal M}^-$ is, in coordinates $(\zeta , \bar\zeta , 
r, u)$,
\begin{equation} \label{2.18}
ds_-^2=2r^2p^{-2}d\zeta\,d\bar\zeta -2\,du\,dr-du^2\ ,
\end{equation}
with
\begin{equation} \label{2.19}
p=\gamma\,\left\{\frac{1}{2}(1+v)\,\zeta\,\bar\zeta +1-v\right\}\ .
\end{equation}
This latter follows from (\ref{2.13}) with the specialisation 
(\ref{2.16}). The line--element of Minkowskian space--time in 
the region $u>0$ to the future of the null--cone ${\cal N}(u=0)$ and 
which we denote by ${\cal M}^+$ is, in coordinates $(\zeta _+,
 \bar\zeta _+, r_+, u_+)$,
\begin{equation} \label{2.20}
ds_+^2=2r_+^2p_+^{-2}d\zeta _+\,d\bar\zeta _+ -2\,du_+\,dr_+-du_+^2\ ,
\end{equation}
with
\begin{equation} \label{2.21}
p_+=\frac{1}{2}\zeta _+\,\bar\zeta _+ +1\ ,
\end{equation}
and $u_+=0$ corresponding to $u=0$. We then attach the two 
halves of Minkowskian space--time ${\cal M}^-$ and 
${\cal M}^+$ on $u=0 
(\Leftrightarrow u_+=0)$ with the {\it matching conditions} 
(cf. (\ref{2.14}))
\begin{equation} \label{2.22}
\zeta _+=\zeta \ ,\qquad \bar \zeta _+=\bar\zeta\ ,\qquad 
r_+=r\,p_+p^{-1}\ .
\end{equation}
Thus in particular the line--elements induced on ${\cal N}$ by 
its embedding in ${\cal M}^-$ and ${\cal M}^+$, calculated from (\ref{2.18}) 
with $u=0$ and from (\ref{2.20}) with $u_+=0$ are the 
same induced line--elements. A more convenient form for (\ref{2.18}) 
for our purposes is obtained by introducing the 
coordinates $\xi , \phi$ via
\begin{equation} \label{2.23}
\xi =\frac{1-\frac{1}{2}\zeta\,\bar\zeta}{1+\frac{1}{2}\zeta\,\bar\zeta}
\ ,\qquad \phi =\frac{1}{2i}\log\left (\frac{\zeta}{
\bar\zeta}\right )\ ,
\end{equation}
which results in (\ref{2.18}) becoming
\begin{equation} \label{2.24}
ds^2_-=k^2\,r^2\left\{\frac{d\xi ^2}{1-\xi ^2}+
(1-\xi ^2)\,d\phi ^2\right\}-2\,du\,dr-du^2\ ,
\end{equation}
with $k^{-1}=\gamma\,(1-v\,\xi )$. Also if $\xi _+, \phi _+$ 
are given by (\ref{2.23}) with $\zeta , \bar\zeta$ replaced 
by $\zeta _+, \bar\zeta _+$ we find that (\ref{2.20}) becomes
\begin{equation} \label{2.25}
ds^2_+=r_+^2\left\{\frac{d\xi _+^2}{1-\xi _+^2}+
(1-\xi _+^2)\,d\phi _+^2\right\}-2\,du_+\,dr_+-du_+^2\ .
\end{equation}
The {\it matching conditions} (\ref{2.22}) now read: on 
$u=0 (u_+=0)$
\begin{equation} \label{2.26}
\xi _+=\xi\ ,\qquad \phi _+=\phi\ ,\qquad r_+=k\,r\ .
\end{equation}

To interpret physically what the geometrical construction 
above is describing we note the following: The line--elements 
(\ref{2.24}) and (\ref{2.25}) are two versions of the 
Minkowski line--element which can be transformed one into 
the other by the coordinate transformation
\begin{equation} \label{3.3}
\xi _+=\frac{\xi-v}{1-v\,\xi}\ ,\qquad \phi _+=\phi \ ,\qquad 
r_+=r\ ,\qquad u_+=u\ .
\end{equation}
Here $v$ is a real parameter with $0<v<1$. The transformation (\ref{3.3}) 
is in fact a Lorentz transformation as it can be shown
if one rewrites the line elements (\ref{2.24})  and (\ref{2.25})
in terms of the rectangular Cartesian coordinates 
and time $\{x, y, z, t\}$ and $\{x_+, y_+, z_+, t_+ \}$.
%related respectively to $\{\xi , \phi , r, u\}$ 
%and $\{\xi _+, \phi _+, r_+, u_+\}$. 
The relations between these
two sets of coordinates and the coordinates $\{\xi , \phi , r, u\}$ 
and $\{\xi _+, \phi _+, r_+, u_+\}$  
are given by
\begin{eqnarray} \label{2.27a}
x &=& r\,k\,\sqrt{1-\xi ^2}\,\cos\phi\ ,\\
y &=& r\,k\,\sqrt{1-\xi ^2}\,\sin\phi\ ,\\
z &=& \gamma\,v\,u\,+r\,k\,\xi\ ,\\
t &=& \gamma\,u+r\,k\ ,
\end{eqnarray}
and 
\begin{eqnarray} \label{2.27b}
x_+ &=& r_+\,\sqrt{1-\xi _+^2}\,\cos\phi _+\ ,\\
y_+ &=& r_+\,\sqrt{1-\xi _+^2}\,\sin\phi _+\ ,\\
z_+ &=& r_+\,\xi _+\ ,\\
t_+ &=& u_++r_+\ .
\end{eqnarray}
In terms of the coordinates  $\{x, y, z, t\}$ and $\{x_+, y_+, z_+, t_+ \}$
the line elements  (\ref{2.24})  and (\ref{2.25}) take the
usual form (\ref{2.1}). 
Furthermore the relationship between $\{x, y, z, t\}$ and 
$\{x_+, y_+, z_+, t_+\}$ corresponding to (\ref{3.3}) is obtained 
directly from substituting (\ref{3.3}) into (2.32)--(2.35) and 
using (2.28)--(2.31). The result is the Lorentz transformation
\begin{equation} \label{3.16}
x_+=x\ ,\qquad y_+=y\ ,\qquad z_+=\gamma\,(z-v\,t)\ ,
\qquad t_+=\gamma\,(t-v\,z)\ ,
\end{equation}
where, as always, $\gamma =(1-v^2)^{-\frac{1}{2}}$.

We now wish to calculate the Maxwell field due to the 
charge $e$ performing the motion described above. To 
this end we first express the line--element of ${\cal M}^-\cup {\cal M}^+$ 
in continuous coordinates $\left\{\Xi, \Phi , R, U\right\}$. 
By this we mean coordinates in which the metric tensor components 
are continuous across ${\cal N}$. In such a coordinate system we 
find that the line--element can be written
\begin{equation} \label{2.27}
ds^2=K^2R^2\left\{\frac{d\Xi ^2}{1-\Xi ^2}+(1-\Xi ^2)\,d\Phi ^2\right\}
-2\,dU\,dR-dU^2\ ,
\end{equation}
with $K^{-1}=\gamma\,\left (1-v\,\Xi\right )$. The coordinate 
ranges are $-1\leq\Xi\leq +1, 0\leq\Phi<2\pi, 0\leq R<+\infty , 
-\infty <U<+\infty$. Here ${\cal N}$ is given 
by $U=0$, ${\cal M}^-$ by $U<0$ and ${\cal M}^+$ by $U>0$. It is interesting to 
note that in these coordinates the metric tensor components are 
independent of $U$. Also when $U>0$, (\ref{2.25}) is transformed to 
(\ref{2.27}) with
\begin{eqnarray} \label{2.28}
r_+ & = & k\,R\,\varphi\ ,\\
u_+ & = & \gamma\,U+k\,(1-\varphi )\,R\ ,\\
\xi _+ & = & \varphi ^{-1}\left (\chi +\Xi\right )\ ,\\
\phi _+& = & \Phi\ ,
\end{eqnarray}
where
\begin{equation} \label{2.32}
\varphi =\left (1+2\,\chi\,\Xi +\chi ^2\right )^{\frac{1}{2}}
\ ,\qquad \chi =\frac{\gamma\,v\,U}{K\,R}\ .
\end{equation}
Clearly when $U<0$, (\ref{2.24}) is transformed into 
(\ref{2.27}) by the identity transformation
\begin{equation} \label{2.33}
r=R\ ,\qquad u=U\ ,\qquad \xi =\Xi\ ,\qquad \phi =\Phi\ .
\end{equation}
We see from (\ref{2.28})--(\ref{2.32}) and (\ref{2.33}) 
that on ${\cal N}(U=0\ \Leftrightarrow u=0\ \Leftrightarrow 
u_+=0)$
\begin{equation} \label{2.34}
r_+=K\,R=k\,r\ ,\qquad \xi _+=\xi\ ,\qquad \phi _+=\phi\ ,
\end{equation}
which agrees with the {\it matching conditions} (\ref{2.26}).
In addition it follows from (\ref{2.27}) that the space--time 
${\cal M}^-\cup {\cal M}^+$ is flat everywhere. In particular the Riemann 
tensor vanishes {\it on} ${\cal N}$ for the {\it matching conditions} 
(\ref{2.34}) and thus ${\cal N}$ is not the history of a light--like signal 
such as an impulsive gravitational wave, a gravitational shock 
wave or a light--like shell of matter. Hence our geometrical construction
has not introduced extra gravitational efffects
and describes a pure electromagnetic phenomenon. This had to be checked as
it is well known that glueing two flat spacetimes on a null hypersurface
can lead  to a Riemann tensor which does not vanish on 
this null hypersurface, see for instance \cite{P}.

The electromagnetic field 
due to the charge $e$ in ${\cal M}^-$ and in ${\cal M}^+$ is the Coulomb 
field. Thus the electromagnetic 4--potential on ${\cal M}^-\cup {\cal M}^+$ is 
given via the 1--form field (in the continuous coordinates 
$\left\{\Xi , \Phi , R, U\right\}$)
\begin{equation} \label{2.35}
A=\frac{e}{r_+}\,(du_++dr_+)\,\vartheta\left (U\right )+\frac{e}{R}\,
\left (dU+dR\right )\left (1-\vartheta\left (U\right )\right )\ 
,\end{equation}
with $r_+, u_+$ given in terms of $\Xi , R, U$ by (\ref{2.28}), (2.29) 
and where $\vartheta\left (U\right )$ is the 
Heaviside step function (equal to zero when $U<0$ and equal to 
unity when $U>0$). Thus 
\begin{equation} \label{2.36}
A=A^+_{{\rm Coul}}\vartheta\left (U\right )+A_{{\rm Coul}}\left (1-
\vartheta\left (U\right )\right )\ ,
\end{equation}
where $A^+_{{\rm Coul}}$ is the Coulomb potential 1--form in 
${\cal M}^+$ due to a charge $e$ with geodesic world--line $r_+=0$ 
and $A_{{\rm Coul}}$ is the Coulomb potential 1--form in ${\cal M}^-$ 
due to a charge $e$ with geodesic world--line $r=0$. We shall 
denote the corresponding Coulomb field 2--forms (the exterior derivatives 
of $A^+_{{\rm Coul}}$ and $A_{{\rm Coul}}$) by $f^+_{{\rm Coul}}$ and 
$f_{{\rm Coul}}$ respectively. Noting from (\ref{2.28}), (2.29) that 
\begin{equation} \label{2.37}
\frac{1}{r_+}(du_++dr_+)=\varphi ^{-1}\left (K\,\gamma\,v\,d\Xi
+\frac{\gamma}{K\,R}\,dU+\frac{1}{R}\,dR\right )\ ,
\end{equation}
we find that the candidate for Maxwell 2--form on ${\cal M}^-\cup {\cal M}^+$ 
is
\begin{equation} \label{2.38}
F=f^+_{{\rm Coul}}\,\vartheta\left (U\right )+f_{{\rm Coul}}\,
\left (1-\vartheta\left (U\right )\right )+e\,K\,\gamma\,v\,
\delta\left (U\right )\,dU\wedge d\Xi\ ,
\end{equation}
where $\delta\left (U\right )$ is the Dirac delta function 
singular on ${\cal N}\left (U=0\right )$. The expressions for 
$f^+_{{\rm Coul}}$ and $f_{{\rm Coul}}$ are given below 
in (\ref{2.57}) and (\ref{2.58}). We remark that in terms 
of the basis 1--forms on ${\cal M}^-\cup {\cal M}^+$,
\begin{eqnarray} \label{2.39}
\vartheta ^1 & = & \frac{K\,R\,d\Xi}{\sqrt{1-\Xi ^2}}\ ,\qquad 
\vartheta ^2=K\,R\,\sqrt{1-\Xi ^2}\,d\Phi\ ,\\
\vartheta ^3 & = & dU\ ,\qquad \vartheta ^4=dR+\frac{1}{2}dU\ ,
\end{eqnarray}
we can write (\ref{2.38}) as
\begin{equation} \label{2.41}
F=f^+_{{\rm Coul}}\,\vartheta\left (U\right )+f_{{\rm Coul}}\,
\left (1-\vartheta\left (U\right )\right )-\frac{e\,\gamma\,v\,
\sqrt{1-\Xi ^2}}{R}\,\delta\left (U\right )\,\vartheta ^1
\wedge \vartheta ^3\ .
\end{equation}
We note that the $\delta$--function part is Petrov 
Type N with degenerate principal null direction given 
via the 1--form $\vartheta ^3$. Thus the principal 
null direction is that of the vector field $\partial /
\partial R$. 

To prove that (\ref{2.41}) is a vacuum Maxwell field on ${\cal M}^-\cup {\cal M}^+$ 
excluding $R=0$, but including the future null--cone $U=0$, 
we first return to (\ref{2.38}) and calculate its Hodge dual
\begin{equation} \label{2.42}
{}^*F={}^*f^+_{{\rm Coul}}\,\vartheta\left (U\right )+{}^*f_{{\rm Coul}}\,
\left (1-\vartheta\left (U\right )\right )+e\,K\,\gamma\,v\,
\delta\left (U\right )\,{}^*\left (dU\wedge d\Xi\right )\ .
\end{equation}
Now since $d{}^*f^+_{{\rm Coul}}=0=d{}^*f_{{\rm Coul}}$, with $d$ 
standing for exterior differentiation, we have
\begin{equation} \label{2.43}
d{}^*F=\delta\left (U\right )\,dU\wedge\left ({}^*f^+_{{\rm Coul}}
-{}^*f_{{\rm Coul}}\right )+e\,\gamma\,v\,d\left [K\,\delta\left (
U\right )\,{}^*\left (dU\wedge d\Xi\right )\right ]\ .
\end{equation} 
Using
\begin{eqnarray} \label{2.44}
{}^*\left (dU\wedge d\Xi\right ) &=& -\left (1-\Xi ^2\right )\,
dU\wedge d\Phi\ ,\\
{}^*\left (dU\wedge dR\right ) &=& -K^2R^2\,d\Xi\wedge d\Phi\ ,\\
{}^*\left (dR\wedge d\Xi\right ) &=& \left (1-\Xi ^2\right )\,dR\wedge d\Phi\ ,
\end{eqnarray}
and the explicit expressions
\begin{eqnarray} \label{2.45}
f^+_{{\rm Coul}}&=&\frac{e\,\gamma}{K\,R^2\varphi ^3}\left\{1-
v\,\Xi +\chi\,\left (\Xi -v\right )\right\}\,dU\wedge dR
+\frac{e\,\chi}{R\,\varphi ^3}\,dR\wedge d\Xi
\nonumber \\
&&+\frac{e}{R\,\varphi ^3}\left\{\chi +\gamma ^2v\,(1-v\,\Xi
)\right\}\,                                                  
%\frac{e\,\gamma}{R\,\varphi ^3}\left\{1-v\,\Xi +2\,\chi\,\Xi -v\,\chi +\chi ^2%\right\}\,
dU\wedge d\Xi\ ,
\end{eqnarray}
and
\begin{equation} \label{2.46}
f_{{\rm Coul}}=\frac{e}{R^2}\,dU\wedge dR\ ,
\end{equation}
we find that
\begin{equation} \label{2.47}
\delta\left (U\right )\,dU\wedge\left ({}^*f^+_{{\rm Coul}}
-{}^*f_{{\rm Coul}}\right )=e\,(K^2-1)\,\delta\left (U\right 
)\,dU\wedge d\Xi\wedge d\Phi\ ,
\end{equation}
and
\begin{equation} \label{2.48}
d\left [K\,\delta\left (U\right )\,{}^*\left (dU\wedge d\Xi\right )
\right ]=\delta\left (U\right )\,\frac{d}{d\Xi}\left\{
\left (1-\Xi ^2\right )\,K\right\}\,dU\wedge d\Xi\wedge 
d\Phi\ .
\end{equation}
Hence (\ref{2.43}) reads
\begin{equation} \label{2.49}
d{}^*F=e\,\delta\left (U\right )\,\left\{K^2-1+\gamma\,v\,
\frac{d}{d\Xi}\left [\left (1-\Xi ^2\right )\,K\right ]\right\}
\,dU\wedge d\Xi\wedge d\Phi\ .
\end{equation}
The right side of this equation vanishes since $K^{-1}=\gamma\,
\left (1-v\,\Xi\right )$. Therefore (\ref{2.41}) is a vacuum Maxwell 
field for all $U$ and for $R>0$. The final (type N) term in 
(\ref{2.41}) thus represents a spherical impulsive electromagnetic 
wave having the future null--cone ${\cal N}\left (U=0\right )$ as history 
in Minkowskian space--time. 

The tetrad defined via the 1--forms (2.39), (2.40) on ${\cal M}^-\cup {\cal M}^+$ 
is a half--null tetrad. The tetrad defined via the 1--forms 
$\left\{\vartheta ^1, \vartheta ^2, \omega ^3, \omega ^4\right \}$ 
with
\begin{equation} \label{2.50}
\omega ^3=dR\ ,\qquad \omega ^4=dU+dR\ ,
\end{equation}
is an orthonormal tetrad. The final term in the electromagnetic 
field (\ref{2.41}) when written on this orthonormal tetrad reads
\begin{equation} \label{2.511}
\hat F=-\frac{e\,\gamma\,v\,\sqrt{1-\Xi ^2}}{R}\,\delta\left (U\right )\,
\left (\vartheta ^1\wedge\omega ^4-\vartheta ^1\wedge\omega ^3\right )\ .
\end{equation}
Thus in the laboratory frame $\hat F$ corresponds to an electric 
3--vector and a magnetic 3--vector given respectively by
\begin{equation} \label{2.51}
{\bf E}={\bf E}_0\,\delta\left (U\right )\ ,\qquad 
{\bf H}={\bf H}_0\,\delta\left (U\right )\ ,
\end{equation}
with
\begin{equation} \label{2.52}
{\bf E}_0=\left ({\cal E}, 0, 0\right )\ ,\qquad  
{\bf H}_0=\left (0, {\cal E}, 0\right )\ ,
\end{equation}
and 
\begin{equation} \label{2.53}
{\cal E}=-\frac{e\,\gamma\,v\,\sqrt{1-\Xi ^2}}{R}\ .
\end{equation}
A measure of the intensity of this electromagnetic wave is 
given by
\begin{equation} \label{2.54}
{\cal I}=\frac{1}{8\pi}\,\left (\left |{\bf E}_0\right |^2+
\left |{\bf H}_0\right |^2\right )=\frac{e^2\gamma ^2\left (1-\Xi ^2\right )}
{4\pi\,R^2}\ .
\end{equation}
A measure of the total intensity, ${\cal I}_{{\rm total}}$, of 
this wave is got by integrating (\ref{2.54}) over the spherical 
wave--front $U=0, R={\rm constant}$. The area element, obtained 
from (\ref{2.27}), is 
\begin{equation} \label{2.55}
dA=K^2R^2d\Xi\,d\Phi\ ,
\end{equation}
with $K^{-1}=\gamma\,\left (1-v\,\Xi\right ),\ -1\leq\Xi\leq +1$ 
and $0\leq\Phi <2\pi$. One readily verifies that the total area 
of the wave--front is $4\pi\,R^2$ and that 
\begin{equation} \label{2.56}
{\cal I}_{{\rm total}}=\frac{e^2}{2}\left\{\frac{1}{v}\,\log\left (
\frac{1+v}{1-v}\right )-2\right\}\ ,
\end{equation}
for $0<v<1$. This is typical of the 3--velocity dependence 
of the total intensity of electromagnetic bremsstrahlung 
(see, for example, Jackson's \cite{J} discussion of beta decay). 
For the charge $e$ above the deceleration from 3--velocity $v$ 
to zero 3--velocity in the laboratory frame is instantaneous 
(at $U=0$) and since (\ref{2.56}) has the characteristic 3--velocity 
dependence of electromagnetic bremsstrahlung we shall refer to 
the radiation described by (\ref{2.511}) as {\it instantaneous 
electromagnetic bremsstrahlung}. It is an important example of an 
impulsive electromagnetic wave because (a) its origin is known 
(it is due to the sudden change in the 4--velocity of the charge) 
and (b) it is free from unphysical directional singularities.

The approach to instantaneous electromagnetic bremsstrahlung given 
in this section leads naturally to the gravitational analogue 
described in section 3. For a treatment of instantaneous electromagnetic 
bremsstrahlung that does not involve the ``cut and paste'' procedure, 
see Appendix A.

\setcounter{equation}{0}
\section{Recoil Effect in Gravitation}\indent
 We consider now the analogous situation for a spherically 
symmetric mass $m$ 
in general relativity to that of the charge $e$ discussed in the 
previous section. We shall first describe our geometrical
model and then physically interpret it. 
Hence in place of the Minkowskian line--
element of ${\cal M}^-$ given by (\ref{2.24}) we take the 
Schwarzschild line--element for ${\cal M}^-$ with a source of 
mass $m$:
\begin{equation} \label{3.1}
ds^2_-=k^2\,r^2\left\{\frac{d\xi ^2}{1-\xi ^2}+
(1-\xi ^2)\,d\phi ^2\right\}-2\,du\,dr-
\left (1-\frac{2\,m}{r}\right )\,du^2\ ,
\end{equation}
with $k^{-1}=\gamma\,(1-v\,\xi )$. Also in place of 
(\ref{2.25}) we take ${\cal M}^+$ to be the Schwarzschild space--
time with source of mass $m_+$:
\begin{equation} \label{3.2}
ds^2_+=r_+^2\left\{\frac{d\xi _+^2}{1-\xi _+^2}+
(1-\xi _+^2)\,d\phi _+^2\right\}-2\,du_+\,dr_+-
\left (1-\frac{2\,m_+}{r_+}\right )\,du_+^2\ .
\end{equation}
For greater generality, and to enable a comparison with 
known results, we have assumed that the rest--mass of the 
Schwarzschild source has changed from $m$ in ${\cal M}^-$ to 
$m_+$ in ${\cal M}^+$.

In (\ref{3.1}) $u={\rm constant}$ are null hypersurfaces 
generated by the integral curves of the vector field 
$\partial /\partial r$ and these null hypersurfaces 
are asymptotically (for large $r$) future null--cones. In 
(\ref{3.2}) $u_+={\rm constant}$ are also null hypersurfaces 
which asymptotically (for large $r_+$) are future null--cones. 
In (\ref{3.1}) we shall take $u\leq 0$ and ${\cal M}^-$ as the region of space--time 
to the past of the null hypersurface ${\cal N}(u=0)$. In (\ref{3.2}) 
we take $u_+\geq 0$, with $u_+=0 \Leftrightarrow u=0$, 
and we take ${\cal M}^+$ to be the region of space--time to the 
future of ${\cal N}$. By analogy with the electromagnetic case 
discussed in section 2 we match ${\cal M}^-$ and ${\cal M}^+$ on ${\cal N}$ 
with the {\it matching conditions} (\ref{2.26}) which, 
by (\ref{3.1}) and (\ref{3.2}), ensure that the line--elements 
induced on ${\cal N}$ by its embedding in ${\cal M}^-$ and ${\cal M}^+$ 
agree.

A physical interpretation of the above geometrical construction 
can be done along the same lines as in section 2
for the electromagnetic case. In particular
one can also perform the same transformation (\ref{3.3})
and make the same change of coordinates  (\ref{2.27a})-(2.31)
and  (\ref{2.27b})-(2.35). Then the line element
(\ref{3.1}) reads
\begin{equation} \label{3.8}
ds^2_-=dx^2+dy^2+dz^2-dt^2+O\left (\frac{m}{{\cal R}}\right )\ ,
\end{equation}
with
\begin{equation} \label{3.9}
{\cal R}=\left\{ x^2+y^2+\gamma ^2(z-v\,t)^2\right\}^{\frac{1}{2}}\ ,
\end{equation}
and for the line element (\ref{3.2}) one gets 
\begin{equation} \label{3.14}
ds^2_+=dx_+^2+dy_+^2+dz_+^2-dt_+^2+O\left (\frac{m_+}{{\cal R}_+}\right )\ ,
\end{equation}
with
\begin{equation} \label{3.15}
{\cal R}_+=\left\{ x_+^2+y_+^2+z_+^2\right\}^{\frac{1}{2}}\ .
\end{equation}
The relationship (\ref{3.16}) between the coordinates
${x,y,z,t}$ and ${x_+,y_+,z_+,t_+}$ immediately show
that ${\cal R}_+={\cal R}$. If the source 
of the gravitational field modelled by the space--
time with line--element (\ref{3.2}) is at rest relative 
to a distant observer using the rectangular Cartesian 
coordinates and time $\{x_+, y_+, z_+, t_+\}$ then the 
source of the gravitational field modelled 
by the space--time with line--element (\ref{3.1}) 
may be considered moving with 3--velocity $v$ in 
the $z_+$--direction relative to this distant observer. 
As a result the physical situation described in the 
opening paragraph of this section is the analogue for a 
mass to that for a charge $e$ described in section 2. 
To a distant observer using the plus coordinates the mass $m$ 
%which may be a black--hole, 
is initially moving rectilinearly with uniform 
3--velocity and is suddenly halted at $u=0(u_+=0)$ and 
experiences a change in its rest--mass. We then 
ask what type of signal exists on the null hypersurface 
${\cal N}$? The theory of 
light--like signals in general relativity  developed
by  Barrab\`es--Israel (BI) \cite{BI}
is tailor--made to answer this question. 

With the space--time ${\cal M}^-$ with line--element (\ref{3.1}) 
attached to the space--time ${\cal M}^+$ with line--element (\ref{3.2}) 
on the null hypersurface ${\cal N}(u=0 \Leftrightarrow u_+=0)$ with the 
{\it matching conditions} (\ref{2.26}) the BI theory enables 
us to calculate, if it exists, the coefficient of $\delta(u)$ 
in the Einstein tensor of ${\cal M}^-\cup {\cal M}^+$. This coefficient, if 
non--zero, is simply related to the surface stress--energy 
tensor of a light--like shell with history ${\cal N}$. The theory also 
enables us to calculate the coefficient of $\delta (u)$ 
in the Weyl tensor of ${\cal M}^-\cup {\cal M}^+$ if it exists. This allows 
us to determine whether or not the light--like signal with history 
${\cal N}$ includes an impulsive gravitational wave \cite{BBH}. For 
the details of the BI technique the reader must consult 
\cite{BI} and further developments are to be found in \cite{BH}. 
We will merely guide the reader through the present application 
of the theory.

The local coordinate system in ${\cal M}^-$ with line--element (\ref{3.1}) 
is denoted $\{x^\mu _-\}=\{\xi , \phi , r, u\}$ while the local 
coordinate system in ${\cal M}^+$ with line--element (\ref{3.2}) is 
denoted $\{x^\mu _+\}=\{\xi _+, \phi _+, r_+, u_+\}$. The equation 
of ${\cal N}$ is $u=0 \Leftrightarrow u_+=0$ and thus we take as normal 
to ${\cal N}$ the null vector field with components $n_\mu$ given via 
the 1--form $n_\mu dx^\mu _{\pm}=-du$. Since we want the physical 
properties of ${\cal N}$ observed by the observer using the 
plus coordinates we take as intrinsic coordinates on 
${\cal N}$, $\{\xi ^a\}=\{\xi _+, \phi _+ , r_+\}$ with $a=1, 2, 3$. 
A set of three linearly independent tangent vector fields to ${\cal N}$ 
is $\left\{e_{(1)}=
\partial /\partial\xi _+, e_{(2)}=\partial /\partial\phi _+, 
e_{(3)}=\partial /\partial r_+\right\}$. The components of these 
vectors on the plus side of ${\cal N}$ are 
$e^\mu _{(a)}|_+=\delta ^\mu _a$. The components of these 
vectors on the minus side of ${\cal N}$ are 
\begin{equation} \label{3.17}
e^\mu _{(a)}|_-=\frac{\partial x^\mu _-}{\partial\xi ^a}\ ,
\end{equation}
with the relation between $\{x^\mu _-\}$ and $\{\xi ^a\}$ 
given by the {\it matching conditions} (\ref{2.26}). Hence 
we find that
\begin{eqnarray} \label{3.18}
e^\mu _{(1)}|_- &=& (1, 0,-r_+\gamma\,v, 0)\ ,\\
e^\mu _{(2)}|_- &=& (0, 1, 0, 0)\ ,\\
e^\mu _{(3)}|_- &=& (0, 0, \gamma\,(1-v\,\xi _+), 0)\ .
\end{eqnarray}
We need a transversal on ${\cal N}$ consisting of a vector field on 
${\cal N}$ which points out of ${\cal N}$. A convenient such (covariant) vector 
expressed in the coordinates $\{x^\mu _+\}$ is ${}^+N_\mu =
(0, 0, 1, \frac{1}{2}-\frac{m_+}{r_+})$. Thus since $n^\mu =\delta ^\mu _3$ 
we have ${}^+N_\mu n^\mu =1$.We next construct the transversal on 
the minus side 
of ${\cal N}$ with covariant components ${}^-N_\mu$. To ensure that 
this is the same vector on the minus side of ${\cal N}$ as ${}^+N_\mu$ 
when viewed on the plus side we require
\begin{equation} \label{3.21}
{}^+N_ \mu\,e^\mu _{(a)}|_+={}^-N_ \mu\,e^\mu _{(a)}|_-\ ,
\qquad {}^+N_\mu {}^+N^\mu ={}^-N_\mu {}^-N^\mu\ .
\end{equation}
The latter scalar product is zero as we have chosen to 
use a null transversal. We find that
\begin{equation} \label{11}
{}^-N_\mu =\left (\frac{r_+v}{1-v\,\xi _+}, 0, \frac{1}
{\gamma\,(1-v\,\xi _+)}, D\right )\ ,
\end{equation}
with
\begin{equation} \label{111}
D=\frac{v^2(1-\xi _+^2)\,\gamma}{2(1-v\,\xi _+)}+
\frac{1}{2\gamma\,(1-v\,\xi _+)}-\frac{m}
{\gamma ^2(1-v\,\xi _+)^2r_+}\ .
\end{equation}
Next the transverse extrinsic curvature on the plus and minus 
sides of ${\cal N}$ is given by
\begin{equation} \label{12}
{}^{\pm}{\cal K}_{ab}=-{}^{\pm}N_\mu\left (\frac{\partial e^\mu 
_{(a)}|_{\pm}}{\partial\xi ^b}+{}^{\pm}\Gamma ^\mu _{\alpha\beta}\,
e^\alpha _{(a)}|_{\pm}e^\beta _{(b)}|_{\pm}\right )\ ,
\end{equation}
where ${}^{\pm}\Gamma ^\mu _{\alpha\beta}$ are the components 
of the Riemannian connection associated with the metric tensor of 
$M^+$ or $M^-$ evaluated on ${\cal N}$. The key quantity we need is 
the jump in the transverse extrinsic curvature across ${\cal N}$ given 
by
\begin{equation} \label{13}
\sigma _{ab}=2\,\left ({}^+{\cal K}_{ab}-{}^-{\cal K}_{ab}\right )
\ .
\end{equation}
This jump is independent of the choice of transversal on ${\cal N}$ 
\cite{BI}. We find that in the present application $\sigma _{ab}=0$ 
except for 
\begin{equation} \label{14}
\sigma _{11}=\frac{2}{1-\xi _+^2}\,\left (m\,k^3-m_+\right )\ ,
\qquad \sigma _{22}=2\,(1-\xi _+^2)\left (m\,k^3-m_+\right )\ ,
\end{equation}
with $k^{-1} = \gamma (1-v \xi)$. 
Now $\sigma _{ab}$ is extended to a 4--tensor field on ${\cal N}$ with 
components $\sigma _{\mu\nu}$ by padding--out with zeros (the 
only requirement on $\sigma _{\mu\nu}$ is $\sigma _{\mu\nu}\,
e^\mu _{(a)}|_{\pm}\,e^\nu _{(b)}|_{\pm}=\sigma _{ab}$). With our 
choice of future--pointing normal to ${\cal N}$ and past--pointing 
transversal, the surface stress--energy tensor components are 
$-S_{\mu\nu}$ with $S_{\mu\nu}$ given by 
\cite{BI}
\begin{equation} \label{15}
16\pi\,S_{\mu\nu}=2\,\sigma _{(\mu}\,n_{\nu )}-\sigma\,n_\mu\,n_\nu 
-\sigma ^{\dagger}g_{\mu\nu}\ ,
\end{equation}
with
\begin{equation} \label{16}
\sigma _\mu =\sigma _{\mu\nu}\,n^\nu\ ,\qquad \sigma ^{\dagger}
=\sigma _\mu\,n^\mu\ ,\qquad \sigma =g^{\mu\nu}\gamma _{\mu\nu}\ .
\end{equation}
In the present case $\sigma _\mu =0$ and thus $\sigma ^{\dagger}=0$ and 
the surface stress--energy tensor takes the form
\begin{equation} \label{17}
-S_{\mu\nu}=\rho\,n_\mu\,n_\nu\ .
\end{equation}
Hence the energy density of the light--like shell measured 
by the distant observer using the plus coordinates is \cite{BI} 
\begin{equation} \label{18}
\rho =\frac{\sigma}{16\pi}=\frac{1}{4\pi\,r_+^2}\,
\left (m\,k^3-m_+\right )\ .
\end{equation}
Thus the null--cone ${\cal N}$ is the history of a light--like shell 
with surface stress--energy given by (\ref{17}). We 
note that $m\,k^3$ is the ``mass aspect"  
in the terminology of Bondi et al.\cite{BBM}, on 
the minus side of ${\cal N}$. 
A calculation of the singular $\delta$--part of the Weyl tensor 
for $M^-\cup M^+$ reveals that it vanishes. Hence there is 
no possibility of the light--like signal with history ${\cal N}$ containing 
an impulsive gravitational wave. We note that $\rho$ is 
a monotonically increasing function of $\xi _+$. Thus on the 
interval $-1\leq\xi _+\leq +1$, $\rho$ is maximum at $\xi _+=+1$ (in the direction of the 
motion) and $\rho$ is minimum at $\xi _+=-1$. This 
is as one would expect. A burst of null matter predominantly 
in the direction of motion is required to halt the mass. 
In this sense the model we 
have constructed here could be thought of as a limiting case of 
a Kinnersley rocket \cite{K} \cite{B}.

By integrating (\ref{18}) over the shell with 
area element $dA_+=r_+^2\,d\xi _+\,d\phi _+$ and with $-1\leq\xi _+\leq +1, 
0\leq\phi _+<2\pi$ we obtain the total energy $E_+$ of the shell 
measured by the distant observer who sees the mass $m$, moving 
rectilinearly with 3--velocity  $v$ in the direction $\xi _+=+1$, 
suddenly halted. Thus
\begin{equation} \label{19}
E_+=\frac{1}{4\pi}\,\int_{0}^{2\pi}d\phi _+\,\int_{-1}^{+1}
\left (m\,k^3-m_+\right )\,d\xi _+\ .
\end{equation}
This results in
\begin{equation} \label{20}
E_+=m\,\gamma -m_+\ .
\end{equation}
So the energy of the light--like shell is the difference in 
the relative masses before and after the 
emission of the light--like shell. If one wants
to exhibit the conservation of energy one can also
interpret (\ref{20}) by saying that, in the reference frame
$(x_+,y_+,z_+,t_+)$ where the mass $m_+$ is at rest,
the energy $m\,\gamma$
of the ingoing mass is transfered into the rest energy $m_+$
plus the energy $E_+$ of the emitted null shell.
 
When $v=0$ $(\gamma =1)$ 
the energy of the shell is the difference in the rest--masses 
(naturally taking $m_+<m$) and this is a well--known result 
\cite{BI} which may also be thought of as a limiting case 
of the Vaidya \cite{V} solution. 
If $v\neq 0$ and $m=m_+$ then
\begin{equation} \label{21}
E_+=m\,(\gamma -1)\ .
\end{equation}
In this case all of the kinetic energy of the mass $m$ 
before stopping is converted into the relativistic shell.

The simplest way of interpreting our solution is that it
represents the field outside a moving spherical body (a star
for instance) which, at a certain moment of retarded time,
suddenly stops as a consequence of some internal process
such as a laser-like nuclear reaction. The emission of a sharp burst
of null matter (photons or neutrinos) predominantly in the forward direction
provides the recoil momentum which is necessary to put the
body at rest. As we are only interested in the outside field
we have not made any assumption about the structure of the
body except that it is spherically symmetric and is initially moving
with constant velocity. If we were to describe the global structure
of spacetime two possibilities would have to be considered.
The first one would correspond to a point--like body, and the maximal 
analytic extension of spacetime would then show that 
the light--like shell emerges from
the white--hole region and propagates radially to future null
infinity. This situation, which appears to be the closest 
analogue of the charged particle described in section 2, is 
of limited physical significance. The second possibility would 
be that the  body 
has some finite radius larger than its gravitational radius. This is
more realistic as it avoids the white-hole region and in particular
the shell starting from the white-hole singularity. Here
only the outside region has to be considered and the light--like shell
is directly emitted from the surface of the body.  
The idea of an extended body remaining rotationally symmetric 
when suddenly decelerated to rest is, of course, a strong idealization. 
More realistically, one would expect the body to undergo some 
deformations and, thereby, to emit gravitational radiation.  
Such deformation effects, treated with the help of approximation methods, 
could be the subject of further studies, using the idealized situation 
considered in this paper as the starting point. 
The time reverse of both of the above possibilities involves impulsive 
acceleration arising from a radially imploding shell of 
null matter (for the first possibility the white--hole is now replaced 
by a black--hole).

It is not surprising that the recoil effect described above 
does not produce an impulsive gravitational wave since the lowest multipole 
of an isolated body that contributes to gravitational radiation 
is the quadrupole and the change of motion that we have considered
here only brings a dipole moment. 
Although the light--like shell above is 
(geometrically) spherically symmetric its energy density (3.20) is 
not, and is concentrated in the forward direction of motion. We have pointed out following (3.20) that in effect it 
is the axial symmetry of the energy density (3.20) which 
explains why the burst of null matter can result in zero 
recoil velocity of the moving source. Impulsive gravitational waves which are 
spherical, in the sense that they have a future null--cone 
history, do exist but they contain line singularities \cite{P} 
and thus are not really spherically symmetric. If a system 
with multipole moments experiences an explosion, such as a 
supernova, which suddenly changes those moments then the 
sudden change in the quadrupole moment is the dominant 
contribution to an asymptotically spherical impulsive 
gravitational wave \cite{BBH}.

\setcounter{equation}{0}
\section{Asymptotic Collision with Back--Scatter}\indent
We illustrate here some interesting calculations that can 
be carried out using the instantaneous electromagnetic 
bremsstrahlung described in section 2 by the final term in 
(\ref{2.41}). As Penrose \cite{P} says:``this retarded 
radiation approximates a purely impulsive electromagnetic 
wave in the far zone, but after an encounter with a 
gravitational plane (impulsive) wave, the electromagnetic 
wave back--scatters and so ceases to satisfy Huygen's 
principle". Since we have in our possession in (\ref{2.41}) 
an analytic expression for the electromagnetic wave, 
we wish to use it to give an analytic description of the 
remainder of the situation envisaged by Penrose and in 
particular to exhibit the back--scattered electromagnetic 
radiation. We therefore first must take the ``plane 
wave limit" of (\ref{2.41}). To do this we make use of 
a device invented by Robinson and Trautman \cite{RT} and 
developed in the context of Li\'enard--Wiechert electromagnetic 
fields by Hogan and Ellis \cite{HE}. We want to find expression 
for a limit of (\ref{2.41}) which captures the idea that we 
are viewing (\ref{2.41}) at a large distance from the 
charge $e$ and over not too large regions of space. This 
can be achieved by first making the coordinate transformation 
\cite{RT}
\begin{eqnarray} \label{4.1}
U &=& \lambda\,\bar U\ ,\qquad R=\gamma\,\lambda ^{-2}+\lambda ^{-1}
\bar V\ ,\\
\Xi &=& \lambda ^2\bar X\ ,\qquad \Phi =\lambda ^2\bar Y\ ,
\end{eqnarray}
where $\lambda$ is a real parameter. If this is applied 
to the line--element (\ref{2.27}) of ${\cal M}^-\cup {\cal M}^+$ and 
then the limit $\lambda\rightarrow 0$ taken, the result is 
the line--element
\begin{equation} \label{4.3}
ds^2=d\bar X^2+d\bar Y^2-2d\bar U\,d\bar V\ .
\end{equation}
If in addition (4.1),(4.2) is applied to the Maxwell field 
(\ref{2.41}) in its form (\ref{2.38}) and at the same time 
the charge $e$ is rescaled according to \cite{HE}
\begin{equation} \label{4.4}
e=\lambda ^{-2}\bar e\ ,
\end{equation}
then for small $\lambda$ we find , using (\ref{2.47}) and 
(\ref{2.48}) that
\begin{equation} \label{4.5}
f^+_{{\rm Coul}}=O\left (\lambda ^2\right )\ ,\qquad 
f_{{\rm Coul}}=O\left (\lambda ^2\right )\ ,
\end{equation}
while
\begin{equation} \label{4.6}
e\,k\,\gamma\,v\,\delta\left (U\right )\,dU\wedge d\Xi =
\bar e\,v\,\delta\left (\bar U\right )\,d\bar U\wedge d\bar X
+O\left (\lambda ^2\right )\ .
\end{equation}
Thus in the limit $\lambda\rightarrow 0$ the electromagnetic 
field (\ref{2.38}) becomes
\begin{equation}\label{4.7}
F=\bar e\,v\,\delta\left (\bar U\right )\,d\bar U\wedge d\bar X\ .
\end{equation}
This is a plane impulsive electromagnetic wave with history 
the null hyperplane $\bar U=0$ in Minkowskian space--time with 
line--element (\ref{4.3}) and with propagation direction 
$\partial /\partial\bar V$ in this space--time. We note that 
evidence of the origin of the wave (\ref{4.7}) in the 
presence of the velocity parameter $v$ has survived the plane wave 
limit.

We now consider the head--on collision of this plane 
wave with a plane impulsive gravitational wave. The 
space-time picture of the process is given in figure 2.

\begin{figure}[!ht]
\center
\input{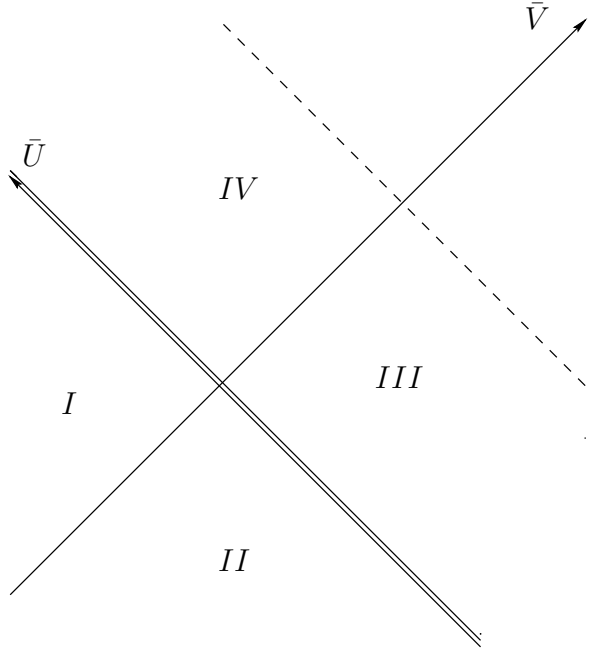}
\caption{\label{2}
The 2--plane $\bar X, \bar Y={\rm constants}$ in 
the space--time having the line--element \cite{P} $\ \  ds^2=(1-p\,\bar V\vartheta (\bar V))^2
d\bar X^2+(1+p\,\bar V\vartheta (\bar V))^2d\bar Y^2
-2\,d\bar U\,d\bar V$. $\ \ \bar U=0, 
\bar V<0$ is the history of the incoming plane electromagnetic 
impulsive wave (\ref{4.7}) while $\bar V=0$ is the history of the 
plane impulsive gravitational wave with which it collides 
head--on. The dotted line is $\bar V = |p|^{-1}$.}
\end{figure}

%\vskip 1truepc
%\centerline{---------------------------------------------}
%\vskip 1truepc
%\centerline{Figure 2}
%\vskip 1truepc
%\centerline{---------------------------------------------}
%\vskip 1truepc\noindent

The half space--time $\bar V<0$ has line--element (\ref{4.3}) 
while the half space--time $\bar V>0$ 
has line--element \cite{P}
\begin{equation} \label{4.8}
ds^2=(1-p\,\bar V)^2d\bar X^2+(1+p\,\bar V)^2d\bar Y^2
-2\,d\bar U\,d\bar V\ ,
\end{equation}
where $p$ is constant. These are both regions of Minkowskian space--
time. The Riemann tensor of the space--time in figure 2 has one 
non--vanishing Newman--Penrose component
\begin{equation} \label{4.9}
\Psi _4=p\,\delta\left (\bar V\right )\ ,
\end{equation}
and is therefore type N in the Petrov classification 
with $\partial /\partial\bar U$ as degenerate principal 
null direction. Thus if $p\neq 0$ then $\bar V=0$ 
in figure 2 is the history of a plane impulsive 
gravitational wave. Also in figure 2, $\bar U=0\ ,\bar V<0$ 
is the history of the plane impulsive electromagnetic 
wave (\ref{4.7}) which collides head--on with the 
gravitational wave. We thus have no Maxwell field within 
regions I, II and III and we solve Maxwell's vacuum 
field equations in region IV assuming no Maxwell 
field on the boundary $\bar V=0, \bar U>0$ and an 
impulsive electromagnetic wave on the boundary $\bar U=0, 
\bar V>0$. In terms of the basis 1--forms in the 
region $\bar V>0$,
\begin{equation} \label{4.10}
\tau ^1=(1-p\,\bar V)\,d\bar X\ ,\qquad 
\tau ^2=(1+p\,\bar V)\,d\bar Y\ ,\qquad 
\tau ^3=d\bar U\ ,\qquad \tau ^4=d\bar V\ ,
\end{equation}
we find that the Maxwell 2--form $F^+$ and its dual ${}^*F^+$ 
in the region $\bar V>0$ is given by
\begin{eqnarray} \label{4.11}
F^+-i{}^*F^+ &=& -\bar e\,v\,(1-p^2\bar V^2)^{-1/2}\,
\delta\left (\bar U\right )\,(\tau ^1+i\tau ^2)\wedge\tau ^3
\nonumber \\
&&+\bar e\,p\,v\,(1-p^2\,\bar V^2)^{-3/2}
\vartheta\left (\bar U\right )\,(\tau ^1-i\tau ^2)\wedge\tau ^4\ .
\end{eqnarray}
Thus the Maxwell 2--form in $\bar V>0$ is 
\begin{eqnarray} \label{4.12}
F^+&=&\bar e\,v\,(1-p\,\bar V)^{1/2}(1+p\,\bar V)^{-1/2}\,
\delta\left (\bar U\right )\,d\bar U\wedge d\bar X\nonumber \\
&&-\bar e\,v\,p\,(1-p\,\bar V)^{-1/2}
(1+p\,\bar V)^{-3/2}\,\vartheta\left (\bar U\right )\,d\bar V\wedge d\bar X\ .
\end{eqnarray}
If the gravitational wave is removed by putting $p=0$ then (\ref{4.12}) 
becomes the plane electromagnetic wave (\ref{4.7}). The {\it first} term 
in (\ref{4.12}) is type N in the Petrov classification with 
degenerate principal null direction $\partial /\partial\bar V$ 
and is an impulsive electromagnetic wave with history $\bar U=0, 
\bar V>0$ and with propagation direction 
$\partial /\partial \bar V$. The geodesic integral curves of 
$\partial /\partial \bar V$ generate $\bar U=0, \bar V>0$ and 
have expansion $p^2\bar V/(1-p^2\bar V^2)$ and shear $p/(1-p^2
\bar V^2)$. Thus $\bar U=0, \bar V>0$ is not a hyper{\it plane} 
and so this impulsive wave is not a plane wave. The {\it second} 
term in (\ref{4.12}), which is non--zero for $\bar U>0$, is 
type N in the Petrov classification with degenerate principal 
null direction $\partial /\partial\bar U$. Thus this term 
describes {\it electromagnetic back--scattered radiation} in region IV
$(\bar U>0, \bar V>0)$ of figure 2 with propagation direction 
$\partial /\partial\bar U$. The integral curves of 
$\partial /\partial\bar U$ are twist--free, expansion--free, 
shear--free null geodesics generating the null hyperplanes 
$\bar V={\rm constant}$. Hence the histories of the 
wave fronts of this back--scattered electromagnetic 
radiation are null hyperplanes. Finally we note that the 
Maxwell field described by (\ref{4.11}) in region $\bar V>0$ 
becomes singular when $\bar V=|p|^{-1}$ where the null geodesics 
tangent to $\partial /\partial\bar V$ are focussed after passing 
through the gravitational wave (see figure 2).

Section 2 above and this section taken together give analytical 
expression to a fascinating space--time diagram (figure 4) in 
\cite{P}. We note that the Maxwell field in region $\bar U>0, 
\bar V>0$ of figure 2, which is given by (\ref{4.11}), not 
only contains the back--scattered radiation but is only 
valid for $0\leq\bar V<|p|^{-1}$.

\noindent
\section*{Acknowledgment}\noindent
We thank Professor W. Israel for helpful discussions. 
This collaboration has been funded by the Minist\`ere des Affaires 
\'Etrang\`eres, D.C.R.I. 220/SUR/R.

\setcounter{equation}{0}
\appendix
\section{Alternative Approach to Electromagnetic Bremsstrahlung}\indent
The rectangular Cartesian coordinates $\left\{X^\mu\right\}$ 
used in section 2 are continuous across the future 
null--cone ${\cal N}$ which is the history of the 
spherical impulsive electromagnetic wave constructed 
there. It may therefore be of some interest to see 
how the wave is constructed in terms of $\left\{X^\mu\right\}$. 
With ${\cal N}$ given by (2.7) the distances $r$ and $r_+$ 
in (2.10) and (2.11) for the 4--velocities (2.16) and (2.17) 
are given by
\begin{equation} \label{1}
r=\gamma\,\left (X^4-v\,X^3\right )\ ,\qquad r_+=X^4\ .
\end{equation}
The jump in the Coulomb potential across ${\cal N}$ is
\begin{equation} \label{2}
\left [A_{{\rm Coul}}^\mu\right ]=e\,\left (\frac{v'^\mu}{r_+}
-\frac{v^\mu}{r}\right )\ .
\end{equation}
Written out explicitly this reads
\begin{eqnarray} \label{3}
\left [A^1_{{\rm Coul}}\right ] &=& \left [A^2_{{\rm Coul}}\right ]
=0\ ,\\
\left [A^3_{{\rm Coul}}\right ] &=& -\frac{e\,v}{X^4-v\,X^3}\ ,\\
\left [A^4_{{\rm Coul}}\right ] &=&-\frac{e\,v\,X^3}{X^4\left (
X^4-v\,X^3\right )}\ .
\end{eqnarray}
Now the 4--potential (2.46) in coordinates $\left\{X^\mu\right\}$ 
is 
\begin{eqnarray} \label{4}
A^\mu &=& A^{+\,\mu} _{{\rm Coul}}\,\vartheta (u)+A^\mu _{{\rm Coul}}\,
(1-\vartheta (u))\ ,\\
&=& e\,\frac{v'^\mu}{r_+}\,\vartheta (u)+ e\,\frac{v^\mu}{r}\,
(1-\vartheta (u))\ ,
\end{eqnarray}
with
\begin{equation} \label{5}
u=\frac{1}{2}\eta _{\mu\nu}X^\mu\,X^\nu\ .
\end{equation}
By (2.7) $u=0$ (with $X^4>0$) is the equation of ${\cal N}$. The 
electromagnetic tensor corresponding to the 4--potential (\ref{4}) 
takes the form
\begin{equation} \label{6}
F^{\mu\nu}=F^{+\,{\mu\nu}}_{{\rm Coul}}\vartheta (u)+
F^{\mu\nu}_{{\rm Coul}}\,(1-\vartheta (u))+\tilde F^{\mu\nu}
\,\delta (u)\ ,
\end{equation}
with
\begin{equation} \label{7}
\tilde F^{\mu\nu}=\left [A^\mu _{{\rm Coul}}\right ]\,u^{,\nu}-
\left [A^\nu _{{\rm Coul}}\right ]\,u^{,\mu}\ .
\end{equation}
To make the connection between the electromagnetic 
wave described in (\ref{6}) and that in section 2 we proceed as follows: 
The delta function part of (\ref{6}) is given equivalently by the 
2--form
\begin{equation} \label{8}
\tilde F=\frac{1}{2}\tilde F_{\mu\nu}\,dX^\mu\wedge dX^\nu 
=\left [A_{{\rm Coul}\,\mu}\right ]\,dX^\mu\wedge du\ .
\end{equation}
By (A.4) and (A.5) we can write 
\begin{equation} \label{9}
\left [A_{{\rm Coul}\,\mu}\right ]\,dX^\mu =-\frac{e\,v}{1-v\,\xi}\,
d\xi =-e\,k\,\gamma\,v\,d\xi\ ,
\end{equation}
with 
\begin{equation} \label{10}
\xi =\frac{X^3}{X^4}\ ,
\end{equation}
and $k^{-1}=\gamma\,(1-v\,\xi )$. Now (\ref{9}) in (\ref{8}) 
shows that (\ref{8}), multiplied by $\delta (u)$, is the spherical impulsive electromagnetic 
wave constructed in section 2 above. Furthermore on ${\cal N}$ 
we have
\begin{equation} \label{11}
X^3=\frac{\xi\,\rho}{\sqrt{1-\xi ^2}}\ ,\qquad X^4=\frac{\rho}{\sqrt{1-\xi ^2}}\ ,
\end{equation}
with $\rho =\sqrt{(X^1)^2+(X^2)^2}$. Putting $X^1=\rho\,\cos\phi\ ,
X^2=\rho\,\sin\phi$ and substituting this and (\ref{11}) into 
(2.1) yields the induced line--element on ${\cal N}$,
\begin{equation} \label{12}
ds^2=\frac{\rho ^2}{(1-\xi ^2)}\,\left\{\frac{d\xi ^2}{1-\xi ^2}
+(1-\xi ^2)\,d\phi ^2\right\}\ .
\end{equation} 
Thus from (\ref{1}) and (\ref{10}) we recover our 
{\it matching condition} $r_+=k\,r$ and, in addition, 
\begin{equation} \label{13}
\rho =r_+\sqrt{1-\xi ^2}\ ,
\end{equation}
and so in (\ref{12}) we have
\begin{equation} \label{14}
\frac{\rho ^2}{(1-\xi ^2)}=r_+^2=k^2r^2\ .
\end{equation}
It is by no means obvious that (\ref{6}) satisfies Maxwell's 
equations. The verification that it does is given in section 2 
above.

\vfill\eject
\noindent
%Figure 1: The future null--cone ${\cal N}(u=0)$ with $QP=r$ and 
%$Q'P=r'$.
%\vfill\eject
%\noindent
%Figure 2: The 2--plane $\bar X, \bar Y={\rm constants}$ in 
%the space--time having the line--element \cite{P} $\ \  ds^2=(1-p\,\bar V\vartheta (\bar V))^2
%d\bar X^2+(1+p\,\bar V\vartheta (\bar V))^2d\bar Y^2
%-2\,d\bar U\,d\bar V$. $\ \ \bar U=0, 
%\bar V<0$ is the history of the incoming plane electromagnetic 
%impulsive wave (4.7) while $\bar V=0$ is the history of the 
%plane impulsive gravitational wave with which it collides 
%head--on.

\end{document}